\documentclass[twocolumn,amsmath,amssymb]{revtex4}

\usepackage{graphicx}
\usepackage{bm}
\usepackage{hyperref}
\usepackage{color}
\usepackage{ulem}

\begin{document}

\preprint{APS/123-QED}

\title{Characterizing the development of sectoral\\ 
Gross Domestic Product composition}

\author{Raphael Lutz}
\affiliation{
Potsdam Institute for Climate Impact Research (PIK),
P.O. Box 60 12 03,
14412 Potsdam,
Germany}

\author{Michael Spies}
\affiliation{
Potsdam Institute for Climate Impact Research (PIK),
P.O. Box 60 12 03,
14412 Potsdam,
Germany}

\author{Dominik E. Reusser}
\affiliation{
Potsdam Institute for Climate Impact Research (PIK),
P.O. Box 60 12 03,
14412 Potsdam,
Germany}

\author{J\"urgen P. Kropp}
\altaffiliation{
University of Potsdam, Dept. of Geo- \& Environmental Sciences, 
14476 Potsdam, Germany}

\author{Diego Rybski}

\email{ca-dr@rybski.de}

\affiliation{
Potsdam Institute for Climate Impact Research (PIK),
P.O. Box 60 12 03,
14412 Potsdam,
Germany}

\date{\today \enspace -- \jobname}

\begin{abstract}
We consider the sectoral composition of a country's GDP, 
i.e.\ the 
partitioning into 
agrarian, industrial, and service sectors. 
Exploring a simple system of differential equations we characterize the 
transfer of GDP shares between the sectors in the course of 
economic development. 
The model fits for the majority of countries providing 
4 country-specific parameters. 
Relating the agrarian with the industrial sector, a data collapse over all 
countries and all years supports the applicability of our approach.
Depending on the parameter ranges, country development 
exhibits different transfer properties. 
Most countries follow 3 of 8 characteristic paths. 
The types are not random but show distinct geographic and 
development patterns. 
\end{abstract}

\maketitle

\section{Introduction}

During its development, human kind has transgressed various stages of
fundamentally different ecological and technological characteristics. In line
with dramatic population growth, an increasing interaction with the biosphere
and a domination of ecosystems took place. During the neolithic revolution,
around 10,000\,BCE, hunter-gatherer societies were progressively replaced by
agrarian ones with far-reaching consequences such as the formation of
settlements. The industrial revolution is considered as the most carving
development affecting all areas of human life and coming along with the
systematic exploration of fossil energy sources.

From an economic point of view, the increasing significance of services is 
understood as an additional level of development. In fact, agrarian, industrial,
and service sectors are commonly denoted primary, secondary, and tertiary,
respectively. However, the production forms do not completely replace each 
other but are complements and economies have more or less contributions 
from each sector.

Current theories and models on sectoral development are largely influenced by
the work of Clark, Fisher, and Fourastie who developed the `three-sector
hypothesis' in the first half of the 20th century, describing development as a
process of shifting economic activities from the primary via the secondary to the
tertiary sector \cite{Clark1940,Fisher1939,Fourastie1949,Kruger2008}. 
Their research was mainly based on observed historical shifts of
workforce between sectors in today's more developed countries. 
Further recently, 
approaches have concentrated on describing the relationship between shifts in sectoral labor
allocation or Gross Domestic Product (GDP) shares and economic development, often focusing on specific
countries or regions \cite{Raiser2004,Echevarria1997,Kongsamut2001}.
Yet, the universality of the three-sector hypothesis has been challenged,
since it does not well represent labor allocation in today's 
developing countries \cite{Timberlake1983,Pandit1989}.
Different from the
historical pathways of industrialized countries,
shifts of labor force from the primary to the secondary
sector have been relatively low. Instead, 
advancement of the tertiary sector appears disproportionate early, which has been related to
excessive urbanization and different structural conditions 
\cite{Timberlake1983,Pandit1989}.

While existing work has mainly focused on modeling patterns observed in the
United States or in Western Europe, applying a similar analysis in a 
universal model does not exist to the best of our knowledge. 
Furthermore, attention
has mainly been given to sectoral resource allocation, e.g.\ labor input,
rather than to economic output, e.g.\ the fractions of GDP.
Thus, the objective
is to develop a parsimonious description of a countries
sectoral composition of GDP, which is also able to capture
the early advancement of the tertiary sector observed in
developing countries.

\section{Model}

We consider a country $c$ and it's 
sectoral GDP composition, where the fractions $a$, $i$, $s$, 
correspond to the agricultural, industrial, and service sector 
contributions, respectively. 
The fractions of the three sectors add up to unity, $a+i+s=1$.

With economic development, i.e.\ increasing GDP/cap, 
the shares of the GDP shift between the sectors.
We assume the transfer occurs according to a system of 
ordinary differential equations
\begin{eqnarray}
\frac{\text{d}a}{\text{d}g}&=& -k_1 a
\label{eq:dgla}\\
\frac{\text{d}i}{\text{d}g}&=& \alpha k_1 a - k_2 i
\label{eq:dgli}\\
\frac{\text{d}s}{\text{d}g}&=& (1-\alpha) k_1 a + k_2 i
\label{eq:dgls} \, ,
\end{eqnarray}

where $g$ is the $\log$ of the GDP/cap 
(the natural logarithm is used in order to compensate for the 
broad distribution of GDP/cap values) 
and $k_{1}, k_{2}, \alpha$ are country-specific parameters 
\footnote[1]{Equations~(\ref{eq:dgla})-(\ref{eq:dgls}) can also be expressed 
with a different set of parameters:\\
$\frac{\text{d}a}{\text{d}g}= -k_{ai}a -k_{as} a$ ,\\
$\frac{\text{d}i}{\text{d}g}= k_{ai} a -k_{is} i$ ,\\
$\frac{\text{d}s}{\text{d}g}= k_{as} a +k_{is} i$ ,\\
with $k_{ai}=\alpha k_1$, $k_{is}=k_2$, and $k_{as}=(1-\alpha)k_1$.}.

\begin{figure}[h]
\begin{centering}
\includegraphics[width=0.6\columnwidth]{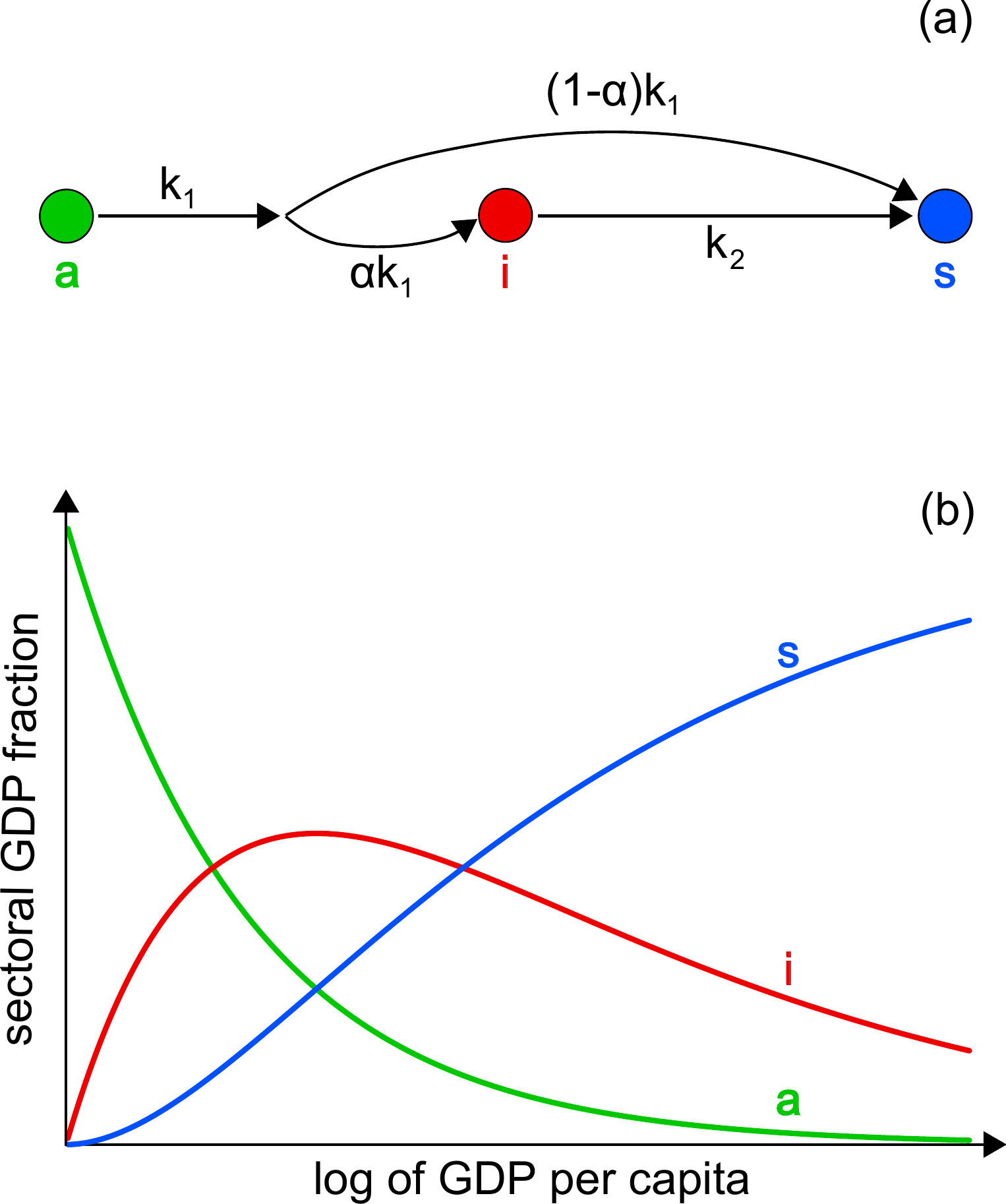}
\caption{Illustration of the transfer model. 
(a) The three sectors, agrarian $a$, industrial $i$, and service $s$, 
are represented by colored circles. 
The arrows indicate possible transfer paths and their parameters as 
defined in Eqs.~(\ref{eq:dgla})-(\ref{eq:dgls}).
(b) Typical evolution of sectoral composition as a function of 
the logarithm of per capita GDP 
(here: $k_1=1$, $k_2=0.5$, $\alpha=1$).}
\label{fig:illustrative}
\end{centering}
\end{figure}

An additional parameter, $g_0$, emerges from the boundary conditions, 
i.e.\ from the value of $g$ where $a=1$, and has the 
character of a shift along the $g$-axis.
Figure~\ref{fig:illustrative} illustrates the model and 
shows schematic trajectories. 
Parameter~$k_1$ determines the transfer from the agrarian sector, 
which, depending on $\alpha$, is split into contributions to the 
industrial or service sectors. 
Moreover, $k_2$ determines the transfer from industrial to service.
E.g.\ for $k_1>0$, $k_2>0$, and $\alpha=1$ transfer takes place 
from~$a$ to~$i$ and continuously from $i$ to $s$, 
leading to monotonously decreasing agrarian and increasing service 
whereas the industrial sector exhibits a maximum 
(Fig.~\ref{fig:illustrative}(b)). 
Except for the trivial case, the model does not have any steady state.

\section{Results}

\begin{table}[h]
\begin{tabular}{|l||p{0.75cm}p{0.75cm}p{0.75cm}|l||l|}
\hline
type & $k_1$ & $k_2$ & $\alpha$ & transfer  & number of \\  
	 &       &       &               & behavior  & countries \\  
\hline\hline
1 & $> 0$ & $> 0$ & $< 1$ & $a\rightarrow i$, $i\rightarrow s$, $a\rightarrow s$ & 59\\
2 & $> 0$ & $> 0$ & $> 1$ & $a\rightarrow i$, $i\leftrightarrows s$, $a \text{ -- } s$  & 25\\
3 & $> 0$ & $< 0$ & $< 1$ & $a\rightarrow i$, $i\leftarrow s$,  $a\rightarrow s$ & 43\\
4 & $> 0$ & $< 0$ & $> 1$ & $a\rightarrow i$, $i\leftarrow s$,  $a \text{ -- } s$  & 0 \\
5 & $< 0$ & $> 0$ & $< 1$ & $a\leftarrow i$,  $i\rightarrow s$, $a\leftarrow s$  & 3 \\
6 & $< 0$ & $> 0$ & $> 1$ & $a\leftarrow i$,  $i\rightarrow s$, $a \text{ -- } s$ & 1 \\
7 & $< 0$ & $< 0$ & $< 1$ & $a\leftarrow i$,  $i\leftarrow s$,  $a\leftarrow s$  & 1 \\
8 & $< 0$ & $< 0$ & $> 1$ & $a\leftarrow i$,  $i\rightleftarrows s$,  $a \text{ -- } s$ & 5 \\
\hline
\end{tabular}
\caption{
Types of sectoral GDP transfer as obtained from the 
fitting parameters of the model Eqs.~(\ref{eq:dgla})-(\ref{eq:dgls}). 
The types are defined according to the sign of the 
parameters $k_1$ and $k_2$ as well as if $0<\alpha<1$ or $\alpha>1$. 
The parameters also specify between which sectors there is a transfer and 
in which direction. 
For $\alpha > 1$ instead of a transfer from the agriculture to the services sector, 
a transfer occurs between industry and services, depending on the values of
$k_1$ and $k_2$ (indicated by a ---).
If $k_1$ and $k_2$ have opposite signs, the flow is in one direction for all 
possible values of $a$, $i$, and $s$. 
If $k_1$ and $k_2$ are positive, flow may occur from $s$ to $i$ for large $a$,
while it is reversed for large $i$ (indicated by $\leftrightarrows$). 
Vice versa for $k_1$ and $k_2$ negative.
Only types $1$+$2$ have a convergent asymptotic behavior, namely
$a\rightarrow 0$, $i\rightarrow 0$, $s\rightarrow 1$ for $g\rightarrow\infty$. 
Most countries belong to the types~1-3. 
The remaining types are 
type~5 (Guinea-Bissau, Madagascar, Vanuatu), 
type~6 (Ivory Coast), 
type~7 (Cameroon), and 
type~8 (Burkina Faso, Morocco, Sudan, Venezuela, South Africa).
The special cases $k_1=0$, $k_2=0$, $\alpha=0$ or 
$\alpha=1$ did not occur.
}
\label{tab:types}
\end{table}

We fit the model Eqs.~(\ref{eq:dgla})-(\ref{eq:dgls}) 
with a two step procedure, using global country-level data
\footnote[2]{Country-level data for fitting the model were obtained from the 
World dataBank provided by the World Bank \cite{Worldbank2012}.\\
$g$: GDP per capita based on purchasing power parity (PPP) in constant 2005 international dollars (World dataBank Series Code NY.GDP.PCAP.PP.KD)\\
$a$: Net output of the agricultural sector as percentage of GDP (World dataBank Series Code NV.AGR.TOTL.ZS)\\
$i$: Net output of the industrial sector as percentage of GDP (World dataBank Series Code NV.IND.TOTL.ZS)\\
$s$: Net output of the services sector as percentage of GDP (World dataBank Series Code NV.SRV.TETC.ZS)\\
The data covers the period 1980-2005 in annual resolution.}.
In the first step, the logarithmic form of Eq.~(\ref{eq:sola}), 
as introduced later, was used to identify $k_1$ and 
an initial value of $g_0$ by using a linear
regression between $\log(a)$ and $g$.
In the second step, $k_2$, $\alpha$, and $g_0$ were estimated by using the 
R-implementation of the shuffled complex evolution algorithm \citep{Duan1993} 
to minimize the sum of the mean squared errors between $a$, $i$, $s$ 
from the model and the corresponding observed values.
To obtain more reasonable fits, we restricted the parameter ranges as follows:
$-5 < k_2 < 5$, $0 < \alpha < 5$, and $1 < g_0 < 15$.

For 176 out of 246 countries the available data was sufficient, 
i.e.\ data on GDP/cap was available for at least 4 years, 
and for 137 countries the fitting worked reasonable 
(we choose $0.1$ as the threshold for the sum of the mean squared error 
between the data and the fits of all sectors). 
Due to an anomalous decline of $g$ after 
dissolution of the Soviet Union, disbandment of the Warsaw Pact, 
and the breakup of Yugoslavia, respectively, 
the data before 1995 has been disregarded, 
in the case of the corresponding countries.
For the same reason, data from Liberia and Mongolia prior to 1995 was omitted.

\begin{figure}[h]
\begin{centering}
\includegraphics[width=\columnwidth]{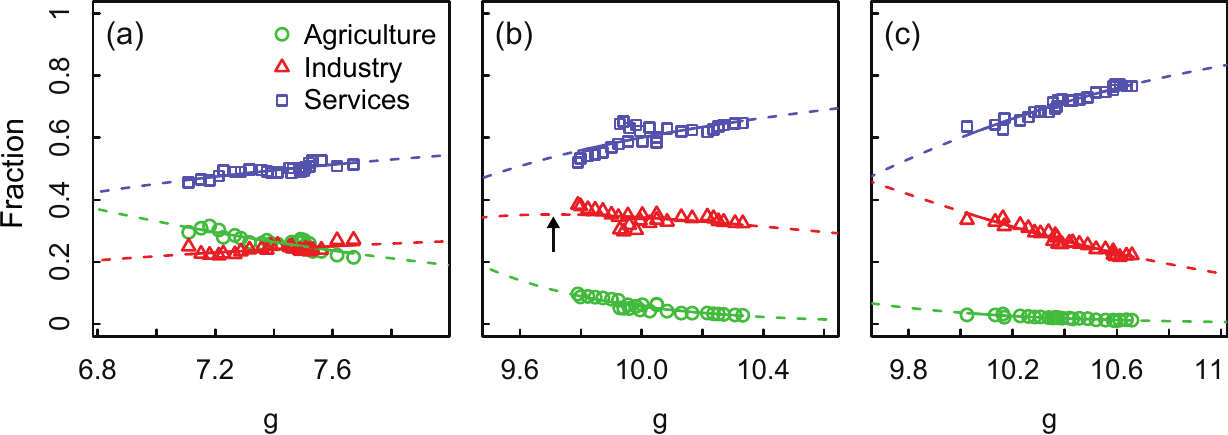}
\caption{Examples of sectoral GDP fractions versus the logarithm of GDP/cap. 
(a) Pakistan, (b) Finland, and (c) the United States. 
Open symbols represent the observed values, solid lines the fitted functions 
according to Eqs.~(\ref{eq:dgla})-(\ref{eq:dgls}), and 
dashed lines extrapolations for illustration.
The fitted parameters are
(a) Pakistan ($k_1=0.56$, $k_2=-0.01$, $\alpha=0.32$, $g_0=8.12$), 
(b) Finland ($k_1=2.29$, $k_2=0.35$, $\alpha=0.50$, $g_0=8.74$), and 
(c) the United States ($k_1=1.76$, $k_2=0.94$, $\alpha=1.27$, $g_0=5.02$).
For Finland, the maximum of the 
industrial sector, $g_{{\rm max}i}\approx 9.7$, 
is indicated by a black arrow in (b).
}
\label{fig:examples}
\end{centering}
\end{figure}

Typical examples for which the model results were accepted 
are depicted in Fig.~\ref{fig:examples} together with 
the obtained fits.
In all three examples the model agrees reasonably with the data. 
In the case of Pakistan, the fraction of industry overtakes the 
fraction of agriculture at $g_{a\times i}\approx 7.5$ ($1,800$\,\$/cap). 
The industrial fraction of Finland reached it's maximum at 
$g_{{\rm max}i}\approx 9.7$ ($16,300$\,\$/cap).
The service sector is the largest and still increasing for these examples. 
In the case of the USA, the agrarian sector has a very low contribution.

Different parameter ranges imply different behavior, 
e.g.\ $k_1<0$ means that the country transfers economic activity 
to agriculture with increasing GDP/cap. 
In total there are two different cases for each parameter leading to eight 
combinations. 
Table~\ref{tab:types} gives an overview of the corresponding types together 
with the transfer behavior, i.e.\ economic transfer from which sector to which, 
and the frequency of each type. 
Almost half of the considered countries belong to type~1, 
the traditional path from the agrarian, via industrial, to the service sector. 
The second most frequent is type~3, which includes a transfer from the service 
to industrial sector. 
Another big group consists of type~2 countries, i.e.\ with transfer from 
agrarian to industry and flows between industry and services depending on
the development. 
All other types are less populated, type~4 does not occur at all. 
The occurrence of types~5-8 might be due to noise in the data. 
Only types~1\&2 and~7\&8 exhibit a maximum of the industrial sector at 
$g_{{\rm max}i}=\frac{\log(k_1/k_2)}{k_1-k_2}+g_0$. 
The examples from Fig.~\ref{fig:examples} are of 
type~3 (Pakistan), type~1 (Finland), and type~2 (USA).

\begin{figure*}[h]
\begin{centering}
\includegraphics[width=0.75\textwidth]{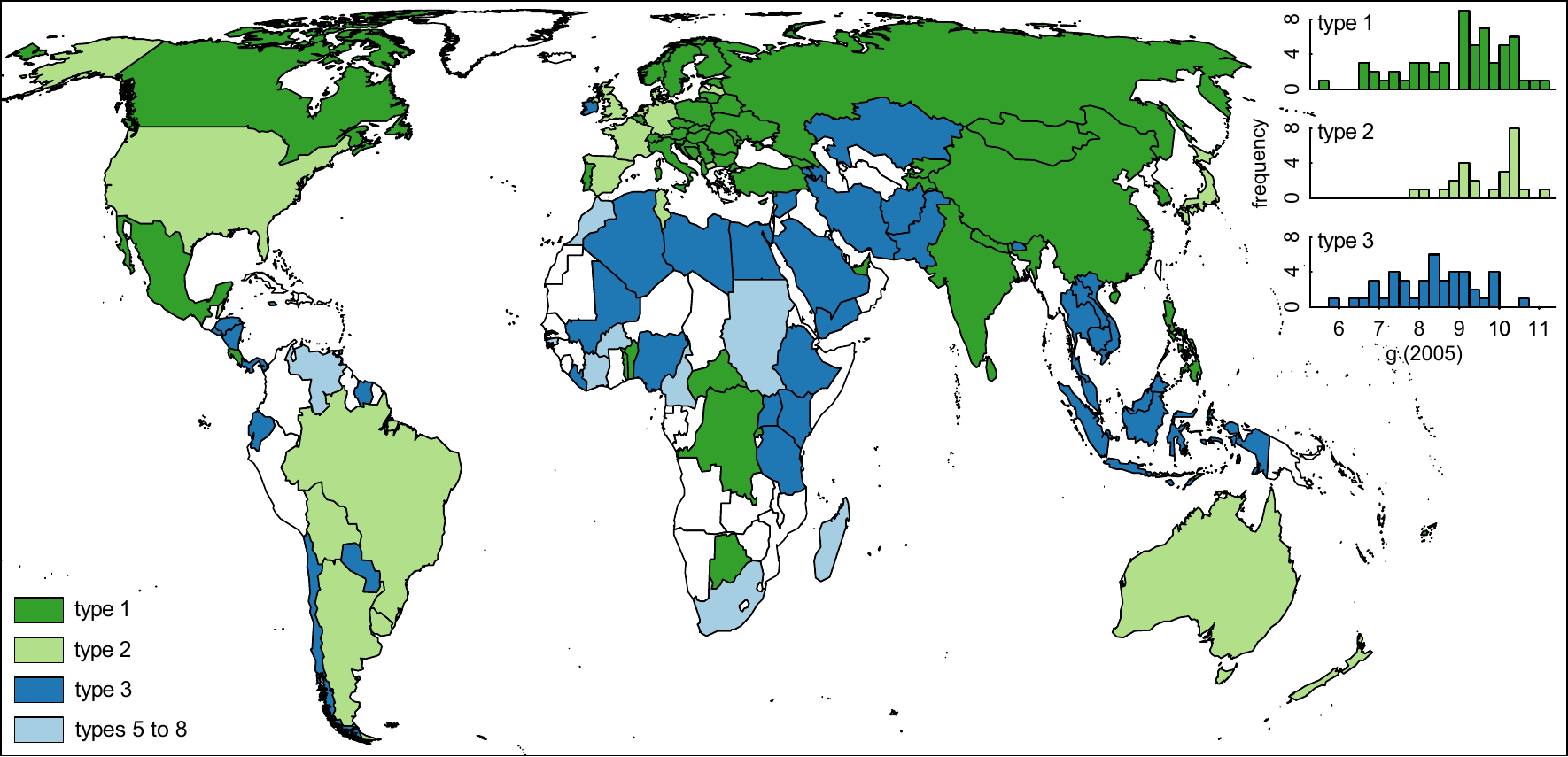}
\caption{Types of sectoral GDP transfer, geographic location, and frequency. 
The resolves the three most common types 1-3 
(Tab.~\ref{tab:types}) and the remaining types 5-8. 
Countries with insufficient data or bad fitting are not colored. 
Countries of the same type tend to be neighbors. 
The inset in the upper right corner shows the frequency of countries 
according to their $g$ (log of GDP/cap) in 2005 for the types 1-3.}
\label{fig:worldfit}
\end{centering}
\end{figure*}

On the world map, Fig.~\ref{fig:worldfit}, one can see 
which country belongs to which type. 
Type~1 consists of big parts of Asia and Eastern Europe, some countries in 
Africa, Canada, and Mexico.
The USA, Brazil, other Southern American countries, Western European countries,
Japan, and Australia belong to type~2. 
Type~3 is mainly found in Africa, Middle East, Central Asia, South-East Asia,
and a few times in Southern America.
A strong regionality can be observed and neighboring countries tend to 
belong to the same types. 
It is apparent that most developed countries belong to either 
type~1 or type~2.
At this stage it is not clear what the decisive factor is and 
further analysis including other economic data could help to 
pinpoint the most relevant influences of the countries 
economic paths.
Methods from network theory have been applied to analyze the economic
productions of countries, indicating that neighboring countries 
instead of diversifying tend to compete over the same markets 
\cite{CaldarelliCGPST2012}.

The inset of Fig.~\ref{fig:worldfit} shows the histograms of 
$g$ (year 2005) for the types 1-3. 
Type~1 and 3 countries are spread over a wide range of GDP/cap, 
whereas there is a tendency of type~1 countries to higher GDP/cap ($g>9$)
compared to the type~3 case.
Type~2 countries generally tend to larger GDP/cap. 
Surprisingly, many high GDP/cap countries belong to type~2 and not to type~1. 
Accordingly, their economic growth follows the traditional path 
$a\rightarrow i\rightarrow s$ but depending on the state of $a$ and $i$, 
the flow between $i \rightarrow s$ may be increased (high $i$), decreased, 
or even reversed (high $a$) (see Tab.~\ref{tab:types}). 
Type~3, which is the second most frequent one, also follows the traditional 
path from the agrarian to the industry and service sector, but comes along with 
a transfer from the service sector to the industry sector, 
$a\rightarrow i\leftarrow s$, $a\rightarrow s$. 
This version seems to be characteristic for many developing economies -- 
although not exclusively.

\section{Data Collapse}

\begin{figure}[h]
\begin{centering}
\includegraphics[width=\columnwidth]{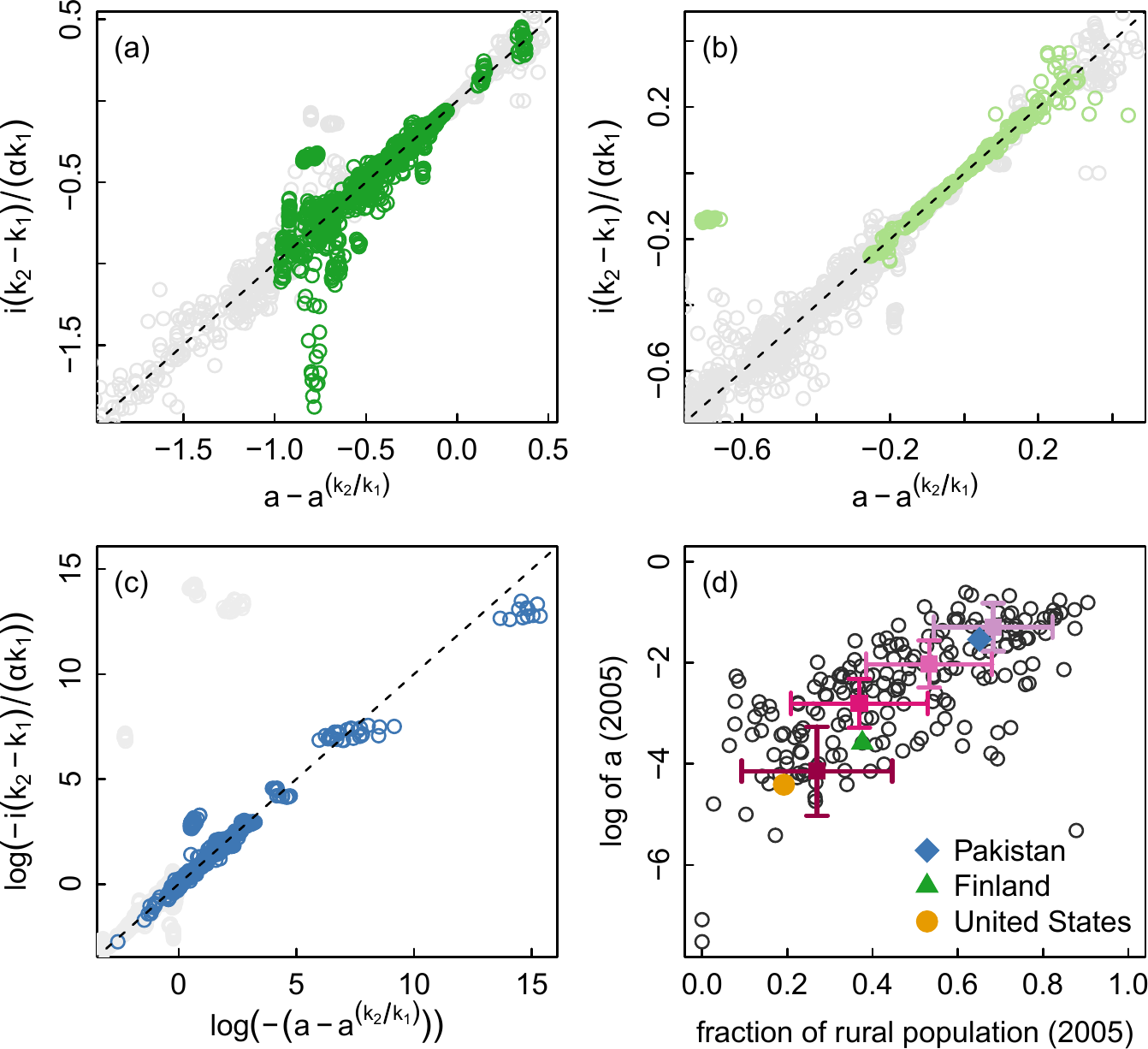}
\caption{Data collapse according to Eq.~(\ref{eq:collapse}) and 
correlations between agrarian sector and rural population. 
Panels (a)-(c) display the transformed values of all countries and all years. 
(a) type~1, (b) type~2, and (c) type~3, see Tab.~\ref{tab:types}.
For clarity, in (c) the signs have been changed and the logarithm taken.
The underlying grey data points show the values of all other types. 
The dashed lines have slope~$1$ and intercept~$0$.
For some countries the fitting does not perform well, in particular
(a) Singapore (at approx. $-0.8$,$-0.35$) and 
Bulgaria (approx. between $-0.8$,$-1.2$ and $-0.8$,$-1.8$), 
(b) Denmark (at approx. $-0.7$,$-0.14$), and 
(c) Saudi Arabia (at approx. $0.6$,$2.9$) and 
the Maldives (at approx. $14.6$,$12.9$).
Panel (d) shows for the year 2005 $\log a$ versus the fraction of 
rural population, based on data from the World dataBank.
The open circles represent the values of all countries with available data 
and exhibit a correlation coefficient of $0.69$.
The examples from Fig.~\ref{fig:examples} are highlighted with 
filled symbols, i.e.\ Pakistan (diamond), Finland (triangle), and 
USA (circle).
The filled squares with error-bars represent averages and standard deviations 
of GDP/cap quartiles, i.e.\ lowest (light, top right) to 
highest (dark, lower left).
}
\label{fig:collapse}
\end{centering}
\end{figure}

In order to test the universality and applicability of the
model, we derive a collapsed representation of all data.
We start from solutions of Eqs.~(\ref{eq:dgla})-(\ref{eq:dgls}) 
\begin{eqnarray}
a(g)&=&{\rm e}^{-k_1 (g-g_0)} \label{eq:sola} \\
i(g)&=&\frac{\alpha k_1}{k_2-k_1}
\left({\rm e}^{-k_1 (g-g_0)}-{\rm e}^{-k_2 (g-g_0)}\right) \label{eq:soli} \\
s(g)&=&1-a(g)-i(g) \label{eq:sols} \, .
\end{eqnarray}
Eliminating $(g-g_0)$ in Eq.~(\ref{eq:sola}) and~(\ref{eq:soli}) 
one obtains a relation between $a(g)$ and $i(g)$
\begin{equation}
\frac{k_2-k_1}{\alpha k_1} i(g) = 
\left(a(g)-a(g)^{k_2/k_1}\right) \, ,
\label{eq:collapse}
\end{equation}
allowing to collapse the data of all countries and all years, 
i.e.\ independent of $g$. 
In Fig.~\ref{fig:collapse}(a)-(c) we plot the transformed data and 
separate the three most frequent types for better visibility into panels. 
The data generally collapses onto the unity diagonal, 
despite few countries where deviations are 
partly due to the fact that values before 1995 have been excluded 
from the fitting (see above) but the values are still 
displayed for completeness.
The collapse suggests 
universality \cite{StanleyHE1999,MalmgrenSCA2009} and 
supports the applicability of the proposed model.

For the set of countries with reasonable fitting, 
some of the parameters are correlated non-linearly, 
i.e.\ $g_0$ and $k_1$ as well as $k_2$ and $\alpha$. 
Thus, by introducing global parameters, the number of country-specific 
ones could be reduced.
Moreover, $k_1$ is weakly correlated with $g$, suggesting that 
-- from the ensemble point of view -- 
Eq.~(\ref{eq:sola}) has rather a log-normal shape, 
which indicates that the system is not ergodic.

We would like to note that since $g$ is the 
logarithm of the GDP/cap, $a$ decreases as a power-law with the GDP/cap.
Studying the asymptotic behavior of the types as defined in 
Tab.~\ref{tab:types}, it turns out, that  types~1 and~2
converge to $s=1$ for $g\rightarrow \infty$. 
The model is confined to specific ranges of $g$, 
e.g.\ $g\le g_0$ for negative parameter $k_1$. 
Thus, it is important to keep in mind that fitting the model 
only characterizes the transfer as it is included in the data. 
This means, the obtained parameters only capture 
the behavior of the past.

\section{Discussion}

Finally, $\log a$ 
is plotted versus the fraction of urban population for the year 2005 
in Fig.~\ref{fig:collapse}(d). 
The two quantities are correlated with a correlation coefficient of $0.69$. 
Despite not being completely linear, the correlations are considerable, 
implying that low agrarian contribution to the economy's GDP comes 
along with less rural population.
In order to visualize the relation to overall economic
output,
Figure~\ref{fig:collapse}(d) also includes averages and
standard deviations 
of those countries belonging to GDP/cap quartiles, i.e.\ the quarter of all 
countries with highest GDP/cap, the second quarter of countries, etc. 
As one can see, with increasing GDP/cap, rurality and agrarian GDP share
decrease. 
In other words, a high degree of urbanization comes along with 
economic development or vice versa.
This can be related to the finding that per capita socio-economic quantities 
such as wages, GDP, number of patents applied, and number of educational 
and research institutions 
increase by an approximate factor of $1.15$ with increasing 
city size \cite{BettencourtW2010}.

However, as Timberlake \cite{Timberlake1983} has pointed out, in the case of
developing countries an ``overurbanization'' with fast growing urban populations
and excessive employment in the service sector can also hinder economic growth. 
Not without reason most developing countries in our model belong to type~3,
where economic growth is associated with a sectoral transfer from service to
industry.

In summary, we propose a system of ordinary differential equations to 
characterize the development of the sectoral GDP composition. 
Despite being very simple and involving only 4 country-specific parameters 
($g_0$ has only the character of shift along $g$), 
the model fits for the majority of the countries in the world.
Relating agrarian and industrial fractions, we collapse the data of 
all countries and all years onto a straight line. 
This could be used as an alternative approach  
to fit the parameters by means of non-linear techniques.

We find that according to the parameter ranges, the countries 
belong to eight different types.
Most countries are found in three of them; 
the members are distinct in geography and state of economic development.
This suggests that countries with low current GDP/cap 
follow a different path from early developed countries.
Our results could indicate a relation between transfer patterns and 
economic development. 
Further analysis of additional socio-economic data could shed light on 
reasons of economic failure or success. 

As with any model, our approach is a strong simplification
of reality. 
Also, we assume that parameters are fixed over time and countries follow a
given development pathway. 
This may be justified by cultural, bio-climatic, and structural conditions, 
which have been consistent over the period of observation. 
On the other hand, a transition between characteristic pathways is possible. 
For workforce distribution of 22 countries from the former
Soviet Union and Central as well as Eastern Europe, 
such a transition analysis has been performed using another simple model
based on the three-sector hypothesis \citep{Raiser2004}. 
A similar transition analysis 
could be an extension of the work presented.  

Since it has been found that countries tend to develop goods 
which are similar to those they currently produce \cite{HidalgoKBH2007} 
and that economically successful countries are extremely diversified 
\cite{TacchellaCCGP2012}, 
it could be also of interest, to extend the analysis to the level of products, 
in order to enable a more detailed analysis. 
Furthermore, the inclusion of a ``quaternary'' 
sector \cite{Kenessey1987,Selstad1990} in our model might provide 
additional insights, but sufficient data is not (yet) available.
In this context we would also like to note that many 
developing countries exhibit an informal service sector which is not included 
in the official figures. 
Similarly, in developed countries the products can be very complex 
so that the separation between industrial and service sector might be fuzzy. 
Accordingly, already the data analyzed in this study is likely to be 
affected by inaccuracies. 


\section*{Acknowledgments}
We thank Torsten Wolpert, Flavio Pinto Siabatto, Xavier Gabaix, 
Boris Prahl, and Lynn Kaack for useful discussions and comments. 
The authors acknowledge the financial support from the Federal Ministry for the
Environment, Nature Conservation and Nuclear Safety of Germany who support this
work within the International Climate Protection Initiative and the Federal
Ministry for Education and Research of Germany who provided support under the
rooftop of the PROGRESS Initiative (grant number \#03IS2191B).


\end{document}